\documentclass[%
 reprint,  
 amsmath,  amssymb,  
 aps,  
]{revtex4-2}

\usepackage{graphicx}% Include figure files
\usepackage{dcolumn}% Align table columns on decimal point
\usepackage{bm}% bold math
\usepackage[acronym]{glossaries}
\usepackage{enumitem}
\usepackage{amsmath}
\usepackage[normalem]{ulem}
\usepackage{color}

\preto{\abstractkeywords}{\nolinenumbers}

\def\al{\alpha}
\def\be{\beta}
\def\ga{\gamma}
\def\de{\delta}
\def\ep{\epsilon}
\def\ze{\zeta}
\def\et{\eta}
\def\th{\theta}
\def\vth{\vartheta}
\def\ka{\kappa}
\def\la{\lambda}
\def\rh{\rho}

\def\si{\sigma}

\def\ta{\tau}

\def\vph{\varphi}

\def\om{\omega}

\def\cL{{\cal L}}

\def\mn{{\mu\nu}}

\def\prt{\partial}
\def\pt#1{\phantom{#1}}

\newcommand{\beq}{\begin{equation}}
\newcommand{\eeq}{\end{equation}}
\newcommand{\bea}{\begin{eqnarray}}
\newcommand{\eea}{\end{eqnarray}}
\newcommand{\rf}[1]{(\ref{#1})}

\def\fr#1#2{{{#1} \over {#2}}}

\def\frac#1#2{{\textstyle{{#1}\over {#2}}}}
\def\rf#1{(\ref{#1})}

\newacronym{gr}{GR}{General Relativity}
\newacronym{sm}{SM}{Standard Model}
\newacronym{cpt}{CPT}{Charge-Parity-Time}
\newacronym{liv}{LIV}{Lorentz invariance violation}
\newacronym{li}{LI}{Lorentz invariance }
\newacronym{gw}{GW}{gravitational wave}
\newacronym{gws}{GWs}{gravitational waves}
\newacronym{lvk}{LVK}{LIGO-Virgo-KAGRA}
\newacronym{sme}{SME}{Standard-Model Extension}
\newacronym{far}{FAR}{false alarm rate}
\newacronym{snr}{SNR}{signal-to-noise ratio}
\newacronym{psd}{PSD}{power spectral density}
\newacronym{ci}{CI}{credible interval}
\newacronym{svd}{SVD}{Singular Value Decomposition}

\begin{document}

\preprint{APS/123-QED}

\title{Search for anisotropic,  birefringent spacetime-symmetry breaking in gravitational wave propagation from GWTC-3}

\author{Le\"ila Haegel}
 \email{l.haegel@ip2i.in2p3.fr}
\affiliation{Université Paris Cité,  CNRS,  Astroparticule et Cosmologie,  F-75013 Paris,  France}

\author{Kellie O'Neal-Ault}
 \email{aultk@my.erau.edu}
 \affiliation{Embry-Riddle Aeronautical University,  Prescott,  AZ,  86301,  USA}
\author{Quentin G. Bailey}
 \email{baileyq@erau.edu}
 \affiliation{Embry-Riddle Aeronautical University,  Prescott,  AZ,  86301,  USA}

\author{Jay D.\ Tasson}
 \affiliation{Carleton College,  Northfield,  MN 55057,  USA}
\author{Malachy Bloom}
\affiliation{Carleton College,  Northfield,  MN 55057,  USA}

\author{Lijing Shao}
\affiliation{Kavli Institute for Astronomy and Astrophysics,  Peking University,  Beijing 100871,  China \\
National Astronomical Observatories,  Chinese Academy of Sciences,  Beijing 100012,  China
}%

\date{\today}

\begin{abstract}
An effective field theory framework,  the Standard-Model Extension,  
is used to investigate the existence of Lorentz and CPT-violating effects during gravitational wave propagation. 
We implement a modified equation for the dispersion of gravitational waves,  that includes isotropic,  anisotropic and birefringent dispersion. 
Using the LIGO-Virgo-KAGRA algorithm library suite,  we perform a joint Bayesian inference of the source parameters and coefficients for spacetime symmetry breaking. 
From a sample of 45 high confidence events selected in the GWTC-3 catalog,  we obtain a maximal bound of $3.19 \times 10^{-15}$~m at 90\% CI for the isotropic coefficient $k_{(V)00}^{(5)}$ when assuming the anisotropic coefficients to be zero.
The combined measurement of all the dispersion parameters yields limits on the order of $10^{-13}$~m for the 16  $k_{(V)ij}^{(5)}$ coefficients.
We study the robustness of our inference by comparing the constraints obtained with different waveform models,  
and find that a lack of physics in the simulated waveform may appear as spacetime symmetry breaking-induced dispersion for a subset of events.  
\end{abstract}

\maketitle

\section{Introduction}
\label{sec:intro}

In the search for a fundamental unified theory of physics,  
it may be imperative to reconsider the axioms underlying \acrfull{gr} and the \acrfull{sm} of particle physics. 
Many theoretical proposals argue for a possible breaking of spacetime symmetries,  
including \acrfull{li} and \acrfull{cpt} symmetry~\cite{ksstring89,  gp99,  chkl01,  ALANKOSTELECKY1991545,  Addazi:2021xuf,  Mariz:2022oib},  
in such a way that it may be detectable in sensitive tests.
The direct detections of \acrfull{gws} reported by the \acrfull{lvk} collaboration provide a new channel to test the rich phenomenology induced by spacetime-symmetry breaking in the gravitation sector~\cite{PhysRevLett.116.061102,  LIGOScientific:2016lio,  LIGOScientific:2019fpa,  Abbott:2020jks,  LIGOScientific:2021djp}. 

The effective field theory referred to as the \acrfull{sme} is a theoretical framework dedicated to derive the observable consequences of spacetime-symmetry breaking that is punctilious and model independent.
The framework is comprised of the action of \acrshort{gr} and the \acrshort{sm} plus all possible terms obtainable from \acrshort{gr} and \acrshort{sm} field operators contracted with coefficients for spacetime-symmetry breaking,  
including local Lorentz,  CPT,  and diffeomorphism breaking terms~\cite{ck97,  ck98,  k04,  bk06,  kt11,  bkx15,  bluhm15,  km16,  km18,  kl21}. 
Extensive constraints have been derived on these terms within the matter sector and in the gravity sector~\cite{datatables},  
the latter having been studied with a wide range of astrophysical probes~\cite{bk06,  km16,  km18,  Xu:2019fyt,  Xu:2021dcw,  nascimento21}.
Existing analysis includes short-range gravity tests~\cite{Long:2014swa,  Shao:2016cjk,  Shao:2018lsx},  gravimetry tests~\cite{Muller:2007es,  Chung:2009rm,  Hohensee:2011wt,  Hohensee:2013hra,  Flowers:2016ctv,  Shao:2017bgz,  Ivanov:2019ouz},  astrophysical tests with pulsars~\cite{Shao:2014oha,  Shao:2018vul,  Wex:2020ald},  solar system planetary tests~\cite{Iorio:2012gr,  Hees:2013iv,  Poncin-Lafitte:2016nqd},  near-Earth tests~\cite{Bourgoin:2016ynf,  Bourgoin:2017fpo,  Bourgoin:2020ckq,  Bars:2019lek},  and tests with \acrshort{gws}~\cite{kt15,  km16,  LIGOScientific:2017zic,  Liu:2020slm,  shao20,  wang21,  Niu:2022yhr}. %cited elsewhere
We complement those searches with further study of the
\acrshort{li} and \acrshort{cpt}-violating effects on propagation of \acrshort{gw}. 
We use dynamical equations for the metric fluctuations derived from the action of the \acrshort{sme},  
and the resulting effects include dispersion,  anisotropy,  and birefringence.

Several tests of \acrshort{gr} have been performed with the \acrshort{gw} events detected by the \acrshort{lvk}~\cite{LIGOScientific:2017zic,  Liu:2020slm,  testGR_GW170817,  shao20,  wang21,  testGR_gwtc3,  wangTestGW2022}.
Some related works focus on parameterizations of the deviations from \acrshort{gr}~\cite{LIGOScientific:2019fpa,  Abbott:2020jks,  Wang_2020,  Wang:2020cub},  including waveform consistency tests,  modification of the \acrshort{gw} generation,  presence of extra polarization modes,  and tests using specific models~\cite{yunes16,  Berti:2018cxi,  Amarilo:2019lfq,  ferrari07,  tso16,  Wang_2020,  Qiao_2019}.
The current searches for \acrshort{li} violation performed by the \acrshort{lvk} collaboration notably rely on a modified dispersion relation that includes isotropic and polarization-independent effects~\cite{testGR_gwtc3,  Mirshekari:2011yq,  Haegel:2021tvb}.
Using the \acrshort{sme} framework,  we extend this phenomenology by measuring the coefficients for \acrshort{li}  and \acrshort{cpt} violation,  including anisotropic and polarization-dependent dispersion.
First estimates of those coefficients have been derived using posterior probabilities released with previous \acrshort{gw} catalog releases,  effectively neglecting the correlations between the parameters describing the source and the spacetime-symmetry breaking coefficients~\cite{shao20,  wang21}. 
In this article,  we present a joint measurement of the source parameters and the coefficients,  alongside studying the robustness of the results we obtain.

Section~\ref{sec:theory} summarizes the derivation of the phenomenology induced by \acrshort{li} and \acrshort{cpt} violation in the \acrshort{sme} framework. 
Section~\ref{sec:method} details the methodological aspects,  including the dataset used for the measurement of the spacetime-symmetry breaking coefficients. 
Section~\ref{sec:results} presents the obtained results,  
as well as a discussion of the impact of the underlying gravitational waveform model and correlations with source parameters.
Section~\ref{sec:conclusion} discusses those results in light of existing studies and future \acrshort{gw} instrument sensitivities.
Theoretical portions of this paper work with natural units,  
where $\hbar=c=1$ and Newton's gravitational constant is $G_N \neq 1$,  
while our data analysis work follows SI units. 
Greek letters are used for spacetime indices while Latin letters for spatial indices. 
We work with the spacetime metric signature $(-,  +,  +,  +)$.

\section{Theoretical derivation of a dispersion relation for gravitational waves}
\label{sec:theory}

We summarize previous derivations in~\cite{km16,  Mewes:2019,  ONeal-Ault:2021uwu},  focusing on gravity-sector terms within the \acrshort{sme} framework. 
The spacetime metric is expanded as fluctuations about the Minkowski metric,  $g_{\mu\nu}=h_{\mu\nu}+\et_{\mu\nu}$,  and we consider up to second order in $h_{\mu\nu}$ for the action,  which is sufficient to characterize propagation effects.
This gives the following action:
\beq
I =\frac{1}{8\ka} 
\int d^4x \,  h_\mn \hat{K}^{(d)\mu\nu\rh\si} h_{\rh\si}.
\label{gravaction}
\eeq
The operator $\hat{K}^{(d)\mu\nu\rh\si}$,  consists of partial derivatives that act on $h_{\mu\nu}$,  
\beq
\hat{K}^{(d)\mu\nu\rh\si} = K^{(d)\mu\nu\rho\si \ep_1 ...\ep_{d-2}}\prt_{\epsilon_1}... \prt_{\ep_{d-2}},  
\label{operators}
\eeq
and $K^{(d)\mu\nu\rho\si \ep_1 ...\ep_{d-2}}$ are  general background coefficients that are considered small,  constant and control the size of any Lorentz or \acrshort{cpt} violation; 

Ensuring linearized gauge symmetry,  i.e.,  $h_{\mu\nu} \rightarrow h_{\mu\nu} + \partial_{\mu}\xi_{\nu}+ \partial_{\nu}\xi_{\mu}$,  and retaining only terms that contribute to the resulting field equations,  we arrive at the following Lagrange density~\cite{km16}:
\bea
        \cL &=& \frac{1}{8\ka} \ep^{\mu\rh\al\ka}\ep^{\nu\si\be\la}\eta_{\ka\la}h_{\mu\nu}\prt_{\al}\prt_{\be}h_{\rh\si} 
        \nonumber\\
        &&+\frac{1}{8\ka} h_{\mu\nu}(\hat{s}^{\mu\rh\nu\si}+\hat{q}^{\mu\rh\nu\si}
        +\hat{k}^{\mu\rh\nu\si})h_{\rh\si}.
         \label{gravlag}
\eea
The first term is the standard GR term written with the totally antisymmetric Levi-Civita tensor density $\ep^{\mu\rh\al\ka}$,  and the remaining terms contain all additional Lorentz invariant and violating terms,  organized into three terms based on symmetry properties: $\hat{s}$ is \acrshort{cpt} even with mass dimension $d \geq 4$; $\hat{q}$ is \acrshort{cpt} odd with mass dimension $d\geq 5$; $\hat{k}$ is \acrshort{cpt} even with mass dimension $d\geq 6$.
Details of these terms including the corresponding Young Tableaux can be found in Table 1 of Ref.~\cite{km16}.
As an example,  for mass dimension $5$,  
\beq
\hat{q}^{\mu\rh\nu\si}=
q^{(5)\mu\rh\ep\nu\ze\si\ka}\prt_\ep \prt_\ze \prt_\ka,  
\label{qhat}
\eeq
where $q^{(5)\mu\rh\ep\nu\ze\si\ka}$ has $60$ independent components.
Note that the gauge symmetry requirement can be relaxed~\cite{km18,  abn21},  but we do not consider such terms here.
The origin of the effective action for $h_\mn$ resulted from explicit symmetry breaking or spontaneous-symmetry breaking as discussed elsewhere~\cite{bk06,  seifert09,  km16,  b21}.

Performing the variation with respect  to $h_\mn$ on the action \eqref{gravlag},  
results in the vacuum field equations,  
\bea
     0 &=& \,  G^{\mu\nu} \,  -[\frac{1}{4}(\hat{s}^{\mu\rh\nu\si}+\hat{s}^{\mu\si\nu\rh})
     +\frac{1}{2}\hat{k}^{\mu\nu\rh\si} \nonumber\\
     &&+\frac{1}{8}(\hat{q}^{\mu\rh\nu\si}
     +\hat{q}^{\nu\rh\mu\si}+\hat{q}^{\mu\si\nu\rh}+\hat{q}^{\nu\si\mu\rh})]\,  h_{\rh\si}.
     \label{eom1}
\eea
Assuming plane wave solutions,  $\bar{h}_\mn = A_\mn e^{-i p_\al x^\al}$,  where $x^\mu$ is spacetime position and $p^\mu = (\om,  \vec p )$ is the four-momentum for the plane wave,  and transforming into momentum space with $\partial_{\alpha}=-i p_{\alpha}$ the dispersion relation can be obtained independently of gauge conditions,  as shown in references~\cite{km09,  km18}.
The dispersion relation for the two propagating modes is given by
\beq
    \omega = |\vec p| \,  
    \left( 1-\zeta^0 \pm |\vec{\ze}| \right),  
    \label{dispEq}
\eeq
where
\beq
|\vec{\ze}|=\sqrt{(\zeta^1)^2 + (\zeta^2)^2 +(\zeta^3)^2}
\eeq
and
\bea
    \ze^0 &=& \frac{1}{4 |{\vec p}|^2} \left(-\hat{s}^{\mu\nu}\,  _{\mu\nu}+\frac{1}{2}\hat{k}^{\mu\nu}\,  _{\mu\nu}\right),  
    \nonumber\\
    (\ze^1)^2+(\ze^2)^2 &=& \frac{1}{8 |{\vec p}|^4}
    \left(\hat{k}^{\mu\nu\rh\si} \hat{k}_{\mu\nu\rh\si}-\hat{k}^{\mu\rh}\,  _{\nu\rh}\,  \hat{k}_{\mu\si}\,  ^{\nu\si}\right. \nonumber\\ && \left. \hspace{75pt} +\frac{1}{8}\hat{k}^{\mu\nu}\,  _{\mu\nu}\,  \hat{k}^{\rh\si}\,  _{\rh\si} \right),  
    \nonumber\\
    (\ze^3)^2 &=&\frac{1}{16|{\vec p}|^4}
    \left(-\frac{1}{2}\hat{q}^{\mu\rh\nu\si}\,  \hat{q}_{\mu\rh\nu\si}
    -\hat{q}^{\mu\nu\rh\si}\,  \hat{q}_{\mu\nu\rh\si}\right. \nonumber\\
    &&\hspace{40pt}\left. +(\hat{q}^{\mu\rh\nu} \,  _{\rh}
    +\hat{q}^{\nu\rh\mu}\,  _{\rh})\hat{q}_{\mu\si\nu}\,  ^{\si}   
    \right).
    \label{zetas}
\eea
We retrieve the \acrshort{gr} case when symmetry-breaking coefficients,  i.e. $\ze^{0}$ and $|\vec{\ze}|$,  vanish.
Note that this result holds at leading order in the coefficients for Lorentz violation,  hence higher modes do not contribute in this perturbative treatment~\cite{km09,  km18}.
Relaxing some of the assumptions in this framework,  allowing for other fields to contribute dynamically to the action,  could result in additional modes~\cite{ybyb14,  Liang:2022hxd}.

\acrshort{gr} predicts two linearly independent polarizations 
for \acrshort{gws} propagating in vacuum,  traveling at the speed of light.
Possible modifications for observable Lorentz and \acrshort{cpt} violating effects from \rf{dispEq} include birefringence,  e.g.,  altered relative travel speeds between the polarizations,  which result from the two possible signs for $|\vec{\ze}|$ in \eqref{dispEq},  requiring a minimum mass dimension $5$. 
Furthermore,  the presence of higher powers of frequency and momentum in the terms above,  indicates beyond \acrshort{gr} dispersion as well.
All of these effects depend on the sky location of the propagating wave,  and thus a breaking of rotational isotropy occurs.

To take into account the sky localization dependence of the source in the detector frame,  it is advantageous to project  the \acrshort{sme} coefficients onto spherical harmonics~\cite{km09},  
\bea
    \ze^0 &=& \sum\limits_{djm} \om^{d-4} \,  
    Y_{jm}(\hat{\textbf{n}})\,  k^{(d)}_{(I)jm},  \label{spherical1}\\
    \ze^1 \mp i\,  \ze^2 &=& \sum\limits_{djm} \omega^{d-4}\,  _{\pm 4}Y_{jm}(\hat{\textbf{n}})\left(k^{(d)}_{(E)jm}\pm i k^{(d)}_{(B)jm}  \right),  
    \label{spherical2}
    \\
    \ze^3 &=& \sum\limits_{djm} \om^{d-4} \,  Y_{jm}(\hat{\textbf{n}})\,  k^{(d)}_{(V)jm},  
    \label{spherical3}
\eea
where  $-j\leq m \leq j$,  the $Y_{jm} (\hat{\textbf{n}}) $ are the standard spherical harmonics while $_{\pm 4}Y_{jm}(\hat{\textbf{n}})$ are spin-weighted spherical harmonics,  and ${\hat n}=-{\hat p}$. 

Expressions for the two linearly independent \acrshort{gw} polarizations,  in the transverse-traceless gauge,  result in a phase shift from the additional symmetry-breaking effects,  
\bea
     h_{(+)} &=& e^{i\de} (\cos \be - i \sin \vth \cos \vph \sin \be )\,  h^{LI}_{(+)} \nonumber\\
     && - e^{i\delta}\sin \be (\cos \vth + i \sin \vth \sin \vph )  \,  h^{LI}_{(\times)} \nonumber\\
     h_{(\times)} &=& e^{i\de} (\cos \be +i \sin \vth \cos \vph \sin \be )\,  h^{LI}_{(\times)} \nonumber\\
     && + e^{i\de}\sin \beta(\cos \vth - i \sin \vth \sin \vph )  \,  h^{LI}_{(+)}.\label{h_LIV}
     \label{plcr}
\eea
where 
\begin{align}
 \de &= \om^{d-3} \tau \zeta^{(d)0},  \nonumber\\
 \be &= \om^{d-3} \tau |\vec{\zeta}^{(d)}|,  \nonumber
 \label{quant1}
\end{align}
and the modified redshift becomes
\beq
\ta = \int_0^z dz \fr{(1+z)^{d-4}}{H(z)}.
\label{redshift}
\eeq
For notational convenience,  the angles in \rf{plcr} are defined by the expressions below, 
\beq
\sin\,  \vartheta=\frac{|\ze^1\mp i\ze^2|}{|\vec{\ze}|},  \pt{30} \cos\vartheta =\frac{\ze^3}{|\vec{\ze}|},  \pt{30} e^{\mp i \varphi}=\frac{\ze^1 \mp i \ze^2}{\sqrt{(\ze^1)^2+(\ze^2)^2}}.
\eeq

One of the key features of spacetime symmetry breaking,  as evidenced in the equations above,  is the breaking of isotropy.
The strength of the \acrshort{li} violation can change with source location~\cite{ONeal-Ault:2021uwu}.
Unless otherwise stated,  the spherical coefficients in~\rf{spherical1}-\rf{spherical3} are expressed in the Sun-Centered Celestial Equatorial reference frame (SCF),  as is standard in the literature~\cite{km02,  datatables},  and allows comparisons with other non-\acrshort{gw} tests in the gravity sector.
Rotations and boosts of the spherical coefficients relative to this frame must be taken into account,  as discussed elsewhere~\cite{kv18}.

\section{Data and parameter inference}
\label{sec:method}

\subsection{Bayesian inference of source and symmetry-breaking parameters}
\label{ssec:infer}

For mass dimension $4$,  \acrshort{li} violation leads to a modification of the \acrshort{gw} group velocity that can be measured with multimessenger signals; constraints on the $\hat{s}$ operator of Eq.~(\ref{gravlag}) have been obtained from the observation of GW170817/GRB170817A~\cite{LIGOScientific:2017zic} comparing light and \acrshort{gw} travel time,  and travel time across the Earth~\cite{Liu:2020slm}.

In this analysis,  we focus on the coefficients for Lorentz and \acrshort{cpt} violation contained in the operator $\hat{q}$ for $d=5$ (see Eq \rf{eom1}),  with the first mass dimension in the action series \rf{gravlag} where \acrshort{gw} dispersion occurs.
We probe the impact of isotropic and anisotropic dispersion,  as well as birefringence,  with a joint estimation of the source parameters and the $16$ {\it a priori} independent $k_{(V)ij}^{(5)}$ coefficients of Eq.~(\ref{spherical3}).
Specifically,  we are considering in this work,  a subset of \rf{plcr},  where $\de=0$.
In this case the remaining coefficients are contained in $\be$.
The expression is lengthy but takes the form~\cite{ONeal-Ault:2021uwu}:
\bea
\beta^{(5)} &=& 
\frac{\omega^2\tau^{(5)}}{2\sqrt{\pi}} \,  \bigl\lvert k^{(5)}_{(V)00}- \sqrt{\frac{3}{2}}\sin \theta \left(e^{i\phi} \,  k^{(5)}_{(V)11}+e^{-i\phi} \,  k^{(5)*}_{(V)11}\right)
\nonumber\\
&&+ \sqrt{3}\cos \theta\,  k^{(5)}_{(V)10}+... \bigr\rvert,  
\label{eq:beta_kv5}
\eea
with the superscript $(5)$ meaning all quantities are evaluated with $d=5$ like equation \rf{redshift}.

Using a Bayesian inference framework,  we compare the strain detected by the \acrshort{lvk} interferometers with a template bank of gravitational waveforms modified as outlined in Eq.~(\ref{plcr}). 
The strain takes the form,  
\beq
S_A = F_{(+)} h_{(+)}+F_{(\times)}h_{(\times)}\label{strainang1},  
\eeq
where $h_{(+,  \times)}$ are the expressions \rf{plcr},  and  $F_{(+,  \times)}$ are the standard detector response functions.
The rotation angles relating different frames,  are included in the expressions for $F_{(+,  \times)}$.
These are defined in the \texttt{LALSuite} software,  including the source frame and the detector frame. 
Again,  the coefficients $k_{(V)ij}^{(5)}$ in \rf{eq:beta_kv5} are left in the SCF.

We use the \texttt{LALSuite} algorithm package,  modifying the 
\texttt{LALSimulation} subpackage to generate dispersed waveforms and performing the parameter estimation with a custom version of \texttt{LALInference}~\cite{Veitch:2014wba}. 
For a single event,  \texttt{LALInference} evaluates the posterior probability with a Markov chain process using the matched-filtered likelihood: 
\bea
   P(d|\vec{\th}_{GR},  \vec{\th}_{SME},  I) = \exp \bigg( \sum_i && - \frac{2 | \tilde{d}_i - \tilde{h}_i (\vec{\th}_{GR},  \vec{\th}_{SME})|^2 }{T  S_n(f_i)}   \nonumber\\
   && - \frac{1}{2} \log \left( \frac{\pi T S_n(f_i)}{2} \right) \bigg)
   \label{eq:llh}
\eea
where $\tilde{h}_i$ is the template signal,  $\tilde{d}_i$ is the interferometer datastream,  $T$ is the duration of the signal,  and $S_n$ the \acrfull{psd} of the detector noise.
Included in the vector set of GR prior parameters $\vec{\theta}_{GR}$,  are the intrinsic parameters describing the binary system (e.g. the masses and spins),  as well as the (extrinsic) astrophysical environment parameters (e.g. the sky location,  distance,  inclination).
The additional parameters $\vec{\theta}_{SME}$,  contain the \acrshort{sme} coefficients $k^{(5)}_{(V)jm}$.

The analysis is performed in the frequency domain,  as detailed in~\cite{ONeal-Ault:2021uwu}. 
%As the \acrshort{psd} is known in the frequency domain,  we use Fourier transformed expressions of $\tilde{h}_i$ and $\tilde{d}_i$ during the inference process.
%Details about the implementation and simulations can be found in~\cite{ONeal-Ault:2021uwu}.
The configurations of the Bayesian inferences,  including the Markov chain algorithms and parameters,  \acrshort{psd} and calibration envelops,  are the same as the ones used by the \acrshort{lvk} collaboration for parameter estimation.
They are retrieved with the \texttt{PESummary} package~\cite{Hoy:2020vys},  selecting the options associated with the \texttt{IMRPhenomPv2} waveform model~\cite{Hannam:2013oca}.

For each \acrshort{gw} event,  we first measure the isotropic dispersion coefficient $|k_{(V)00}^{(5)}|$,  by taking a limiting case of \rf{eq:beta_kv5} where we temporarily ignore the sky angle dependence and assume a flat prior on the \acrshort{sme} coefficient.
We then combine individual posterior probability densities to obtain a measurement of the 16 anisotropic coefficients $k_{(V)ij}^{(5)}$ while taking into account the source sky localization via the full expression in \rf{eq:beta_kv5}.
We perform the combination by interpreting the sampled parameter as the linear combination: 
\beq
\vec{K} = \boldsymbol{Y} \cdot \vec{k}_{(V)ij}^{(5)}
\label{klincomb},  
\eeq
where $\vec{K}$ is the vector of N=45 posteriors of $k_{(V)00}^{(5)} / Y_{00}$,  $\boldsymbol{Y}$ is the matrix of spherical harmonics $Y(\theta, \phi)$  with $\theta$ and $\phi$ the sky coordinates,  and $\vec{k_{(V)ij}^{(5)}}$ corresponds to the n=16 \acrshort{sme} coefficients.
To invert Eq.~(\ref{klincomb}) while the dimensions of the vectors $\vec{K}$ and $ \vec{k_{(V)ij}^{(5)}}$ are unequal,the  $\boldsymbol{Y} $ matrix of size $N \times n$ can be decomposed with the \acrfull{svd} method: 
\beq
\vec{K} = \boldsymbol{U} \boldsymbol{\Sigma} \boldsymbol{V^T} \cdot \vec{k}_{(V)ij}^{(5)}
\label{klinsvd},  
\eeq
where $ \boldsymbol{U}$ is a $N \times n$ matrix of the $ \boldsymbol{Y Y^T }$ orthonormal eigenvectors,   $ \boldsymbol{\Sigma}$ is a $n \times n$ diagonal matrix of the square root of the $ \boldsymbol{Y^T Y}$ eigenvalues,  and $ \boldsymbol{V^T}$ is the transpose of the $n \times n$ matrix of the  $ \boldsymbol{Y^T Y}$ orthonormal eigenvectors.
The  $ \boldsymbol{\Sigma}$ elements are the singular values $\sigma_{0..n}$, chosen to be in decreasing order : $\sigma_1 > \sigma_2 > ... > \sigma_n$.
Inverting Eq.~(\ref{klinsvd}) results in: 
\beq
\vec{k} = \boldsymbol{V} \boldsymbol{\Sigma^{-1}} \boldsymbol{U^{T}} \vec{K}
\label{ksvd},  
\eeq
which is equivalent to performing the linear least square minimization of $ ||\vec{K} - \boldsymbol{Y} \cdot \vec{k}_{(V)ij}^{(5)}||_2 $,  as done in the mesurement of  $\vec{k}_{(V)ij}^{(5)}$ without joint inference of the source parameters~\cite{golub1971singular, shao20}.
We highlight that the results obtained with this method may differ from other type of multi-parameter inference, such as inferring the coefficients from multiple events at the time (currently infeasible due to the long sampling time it requires) or using Bayesian hierarchical inference. 
We chose this method motivated by two aspects: (i) to compare our results with previous estimates, highlighting the differences due to the joint inference of source and symmetry breaking parameters, and (ii) realising that due to the dimensionality and ordering of the $ \boldsymbol{\Sigma}$ matrix, this method effectively put more weight on the 16 events performing the best combination of $ \vec{k_{(V)ij}^{(5)}}$ parameters, effectively leading to an ``optimal" estimate. 

%We invert Eq.~\ref{klincomb} using the \acrfull{svd} method,  for which: 
%\beq
%\vec{k} = \boldsymbol{V} \boldsymbol{\Sigma^{\dagger}} \boldsymbol{U^{T}} \vec{K}
%\label{ksvd},  
%\eeq
%where the $\boldsymbol{V},  \boldsymbol{\Sigma^{\dagger}}$ and $\boldsymbol{U^{T}}$ matrices are the \acrshort{svd} factorisation of $\boldsymbol{Y}$.

\subsection{Dataset from the GWTC-3 catalog}
\label{ssec:data}

We perform our analysis on the events detected during the three first observing runs,  corresponding to the cumulative catalog GWTC-3~\cite{LIGOScientific:2021djp}.
All the \acrshort{gw} detections originate from the coalescence of binary systems of black holes and/or neutron stars,  and the catalog reports 90 events with a probability of astrophysical origin larger than 50\%. 
The study of \acrshort{gw} residual shows that after subtraction of the best-fitted waveforms assuming \acrshort{gr},  the leftover signals are consistent with noise,  indicating that deviations from \acrshort{gr} are higher order terms inducing small modification of the signal morphology~\cite{testGR_gwtc3}.
Therefore,  low-sensitivity events are unsuited for tests of \acrshort{gr} as they may lead to false apparent deviations due to transient noise or incomplete modelling of the gravitational waveform~\cite{Kwok:2021zny,  Moore:2021eok}.
To prevent such undesirable features,  we add the following requirements: (i) the \acrfull{far} must be lower than $10^{-3} \ \text{year}^{-1}$,  in order to only use high-confidence signals; and (ii) the event must have been selected by the \acrshort{lvk} to test the modified dispersion relation,  as while we use a different theoretical framework and phenomenology,  we are also performing a measurement of \acrshort{gw} dispersion. 
The final selection contains 45 events (10 events first reported in GWTC-1,  23 events in GWTC-2,  and 12 events in GWTC-3),  with \acrfull{snr} comprised within [9.2; 26.8] and luminosity distances within $[0.32,  4.42]$\,  Gpc. 

\section{Measuring spacetime symmetry breaking parameters}
\label{sec:results}

\subsection{Constraints with GWTC-3}
\label{ssec:constraints}

\begin{figure}[h!]
\includegraphics[width=\linewidth]{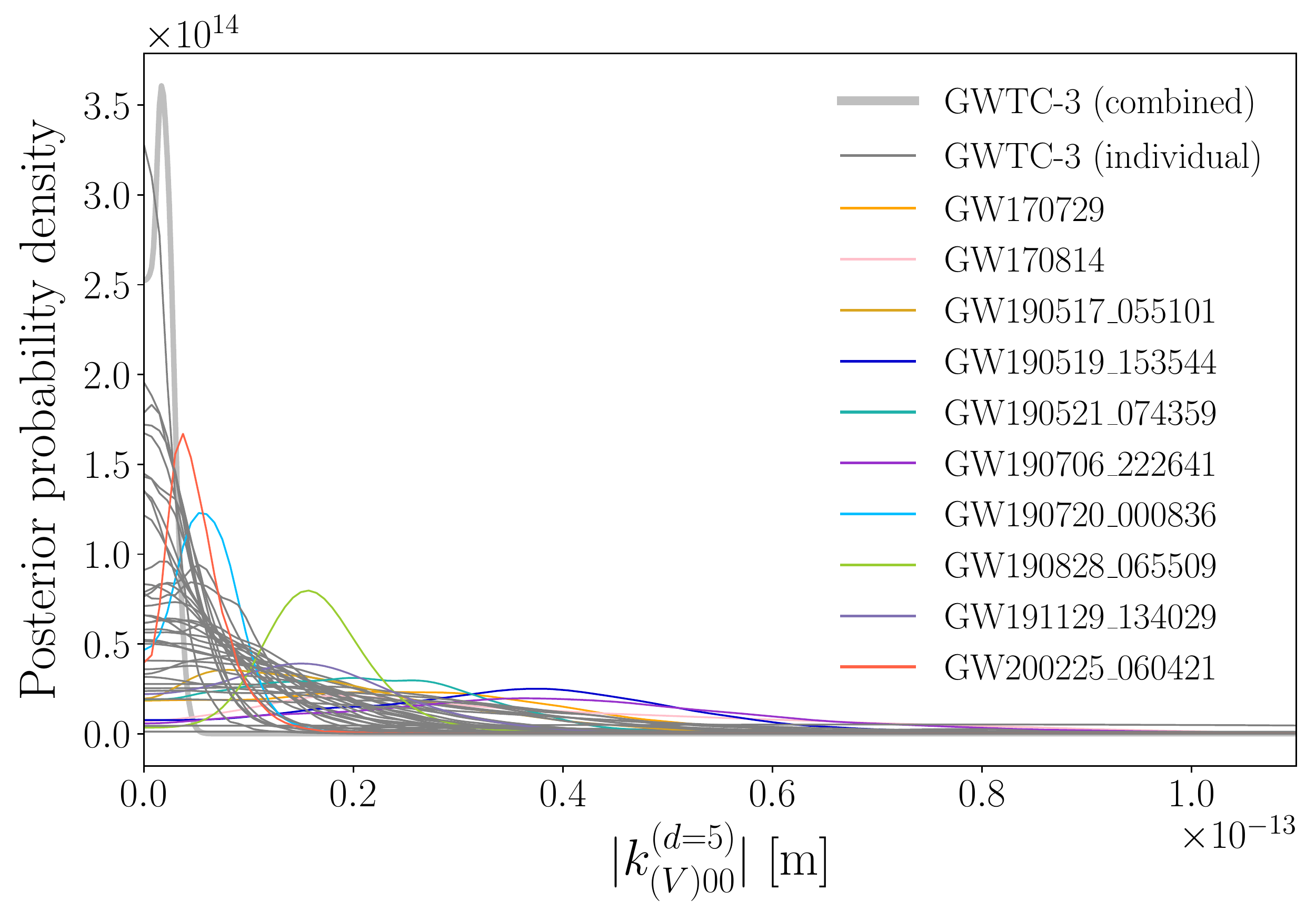}
\caption{\label{fig:kv500} Posterior probability on the isotropic dispersion coefficient $|k_{(V)00}^{(5)}|$  from individual events. The events in color presents a 68.3\% \acrshort{ci} not compatible with the \acrshort{gr} case of $|k_{(V)00}^{(5)}| = 0$,  while the events in grey (\textit{individual}) are compatible.  The thick grey line (\textit{combined}) is the joint constraint when combining all 45 posterior probability densities}. 
\end{figure}

The marginalised posterior distribution of the isotropic dispersion coefficient $|k_{(V)00}^{(5)}|$ is obtained for all the events described on Sec.~\ref{ssec:data}. %ssec:data}. 
As shown in Fig.~\ref{fig:kv500},  most events are compatible with a zero value of $|k_{(V)00}^{(5)}|$,  corresponding to the \acrshort{gr} case. 
The 68.3\% upper bounds range between $\mathcal{O}(10^{-14})$ and $\mathcal{O}(10^{-13})$ according to the event,  with 10 events presenting a 68.3\% \acrlong{ci} not compatible with \acrshort{gr}.
Only one event,  \texttt{GW190828\_065509},  is not compatible with \acrshort{gr} at 90\% \acrshort{ci}.
There have not been any transient noise (or ``glitch") requiring data quality mitigation recorded at the same time as the \acrshort{gw} signals leading to a deviation, from \acrshort{gr}, indicating that instrumental or environmental artifacts are unlikely to be the cause.

The combined constraint from all events on $|k_{(V)00}^{(5)}|$ is $3.19 \cdot 10^{-15}$ m at 90\% \acrshort{ci}.
At 68.3\% \acrshort{ci},  the combined bound is $ 5.62 \cdot 10^{-16} < |k_{(V)00}^{(5)}|< 2.81 \cdot 10^{-15}$ m. 
This deviation from \acrshort{gr} is driven by the events  \texttt{GW190720\_000836},  \texttt{GW190828\_065509},  \texttt{GW200225\_060421},  and is alleviated when removing the three posteriors from the combination.

\begin{figure*}
\includegraphics[width=.9\textwidth]{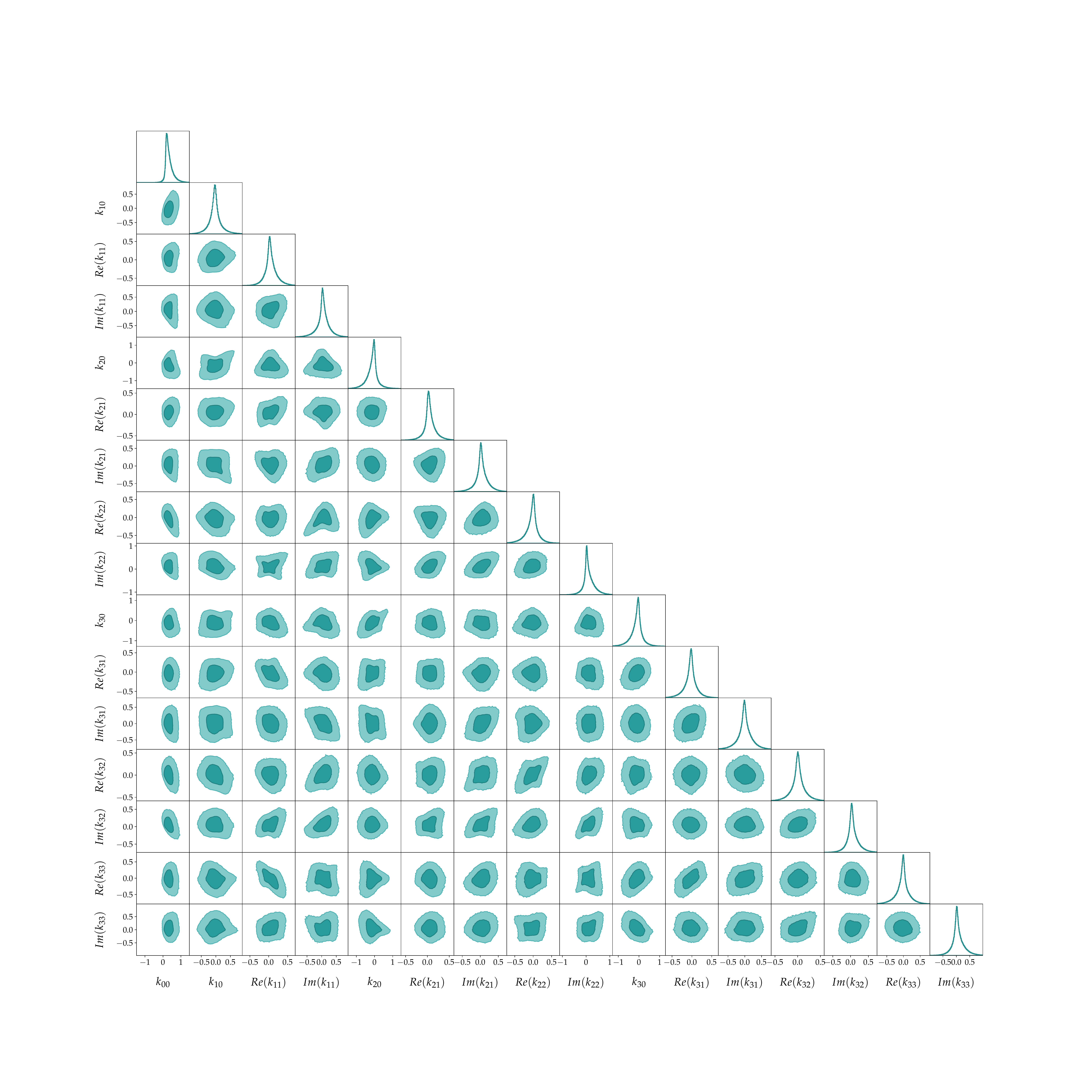}
\caption{\label{fig:kv5ij} Posterior probability of the $k_{(V)ij}^{(5)}$ coefficients (in $10^{-12}$  m).  For the 2-dimensional distribution,  dark blue are the 68.3\% credible intervals and light blue the 90\% credible intervals. }
\end{figure*}

Using the fact that we have more individual events than coefficients,  we perform a sky-localisation dependent analysis to extract the 16 $k_{(V)ij}^{(5)}$ coefficients from the $|k_{(V)00}^{(5)}|$ posteriors,  separating the anisotropic coefficients into real and imaginary components. 
Applying the  \acrshort{svd} inverting method described on Sec.~\ref{ssec:infer},  we obtain the posterior probabilities on the joint estimates shown in Fig.~\ref{fig:kv5ij} alongside the correlations between the parameters. 
We note that during the combination of the posteriors from multiple events,  information from the sky localisation distributions are transferred to the joint posteriors as explicited in Eq~\ref{klincomb}.
The dimensionality reduction of the SVD method not being suitable to perform the common procedure of variable transform that would alleviate the effect of the non-flat prior,  we observed the impact of the angular dependence by performing event-per-event single transformation of one parameter into another (e.g. $k_{(V)00}^{(5)}$ to $k_{(V)10}^{(5)}$, assuming only one non-zero parameter at the time). 
Comparing posterior probability densities on $k_{(V)ij}^{(5)}$ for single events with and without a flat prior in $k_{(V)ij}^{(5)}$, we noticed that the distributions were very similar and therefore the  impact of the sky localisation prior  negligible. 
The bounds on the spacetime symmetry breaking coefficients are extracted from the marginalised 1-dimensional posterior probability distributions displayed diagonally in Fig.~\ref{fig:kv5ij},  and summarised in Table~\ref{tab:kij_svd}.
All the anisotropic coefficients are compatible with the \acrshort{gr} case. 
The isotropic coefficient $k_{(V)00}^{(5)}$ presents a deviation towards values superior to 0,  driven by the events presenting a deviation in Fig.~\ref{fig:kv500}. 
When removing those 10 events,  the deviation from \acrshort{gr} does not appear anymore.
The joint estimation is however less constraining than the individual one,  as the bounds are three orders of magnitude larger than the combined constraint of individual events assuming the anisotropic coefficients to be zero. 

\begin{table}[h!]
\setlength{\tabcolsep}{12pt}
\centering
\begin{tabular}{ c c c c c}
\hline \hline
90\% & 68.3\%  & $k_{(V)ij}^{(5)}$  & 68.3\%  & 90\%  \\ 
lower & lower &  coefficient & upper & upper \\ 
  \hline \hline
      0.51  &        1.21  &  $k_{00}$  &        4.38  &        7.37  \\
     -4.54  &       -2.13  &  $k_{10}$  &        1.19  &        3.91  \\
     -2.30  &       -1.00  &  $Re(k_{11})$  &        1.73  &        3.39  \\
     -3.64  &       -1.21  &  $Im(k_{11})$  &        2.35  &        4.45  \\
     -7.40  &       -3.75  &  $k_{20}$  &        1.10  &        3.78  \\
     -1.75  &       -0.61  &  $Re(k_{21})$  &        1.43  &        3.02  \\
     -2.77  &       -1.16  &  $Im(k_{21})$  &        1.71  &        3.67  \\
     -3.58  &       -1.72  &  $Re(k_{22})$  &        1.02  &        2.55  \\
     -2.49  &       -0.96  &  $Im(k_{22})$  &        2.80  &        5.58  \\
     -6.40  &       -3.31  &  $k_{30}$  &        1.17  &        3.57  \\
     -3.34  &       -1.65  &  $Re(k_{31})$  &        0.98  &        2.48  \\
     -3.90  &       -1.92  &  $Im(k_{31})$  &        1.75  &        3.87  \\
     -2.76  &       -1.23  &  $Re(k_{32})$  &        1.34  &        2.87  \\
     -2.26  &       -0.90  &  $Im(k_{32})$  &        1.82  &        3.60  \\
     -3.95  &       -1.95  &  $Re(k_{33})$  &        1.28  &        3.18  \\
     -3.22  &       -1.35  &  $Im(k_{33})$  &        2.25  &        4.78  \\
  \hline \hline
\end{tabular}
\caption{Credible intervals on the $k_{(V)ij}^{(5)}$ coefficients (in $10^{-13}$  m),  determined from the marginalised posterior probability distributions estimated with the joint estimation of the 16 $k_{(V)ij}^{(5)}$  coefficients shown in diagonal in  Fig.~\ref{fig:kv5ij}.}
\label{tab:kij_svd}
\end{table}

\subsection{Robustness tests}
\label{ssec:robust}

The events shown in color in Fig.~\ref{fig:kv500} present a 68.3\% \acrshort{ci} not compatible with the \acrshort{gr} case of $|k_{(V)00}^{(5)}| = 0$ m.
We have surveyed the results presented by the \acrshort{lvk} in their articles summarising several tests of \acrshort{gr} to look for other pathological behaviour from those events~\cite{LIGOScientific:2019fpa,  Abbott:2020jks,  testGR_gwtc3}. 
We find two events from O2 (\texttt{GW170729},  \texttt{GW170814}) and two from O3 (\texttt{GW190828\_065509},  \texttt{GW200225\_060421}) that have been shown to drive a bias in the estimation of the modified dispersion relation parameters. 
Three O3 events (\texttt{GW190519\_153544},  \texttt{GW190521\_074359},  and  \texttt{GW190828\_065509}) present deviations in parameterised post-Newtonian tests,  while one O3 event (\texttt{GW200225\_060421}) presents poor score in residual tests. 
Those features indicate a lack of new physics in the model used to generate the \acrshort{gr} parts $h_{(+)}^{LI}$ and $h_{(\times)}^{LI}$ of Eq.~(\ref{plcr}),  that may originate from the lack of dynamical effects of features of a new theory.
The mismodeling or lack of modeling of dynamical phenomena,  such as an approximation of precession or assuming circular orbits,  can impact the estimation of beyond-\acrshort{gr} parameters~\cite{Moore:2021eok}.

\begin{figure}[h!]
\includegraphics[width=\linewidth]{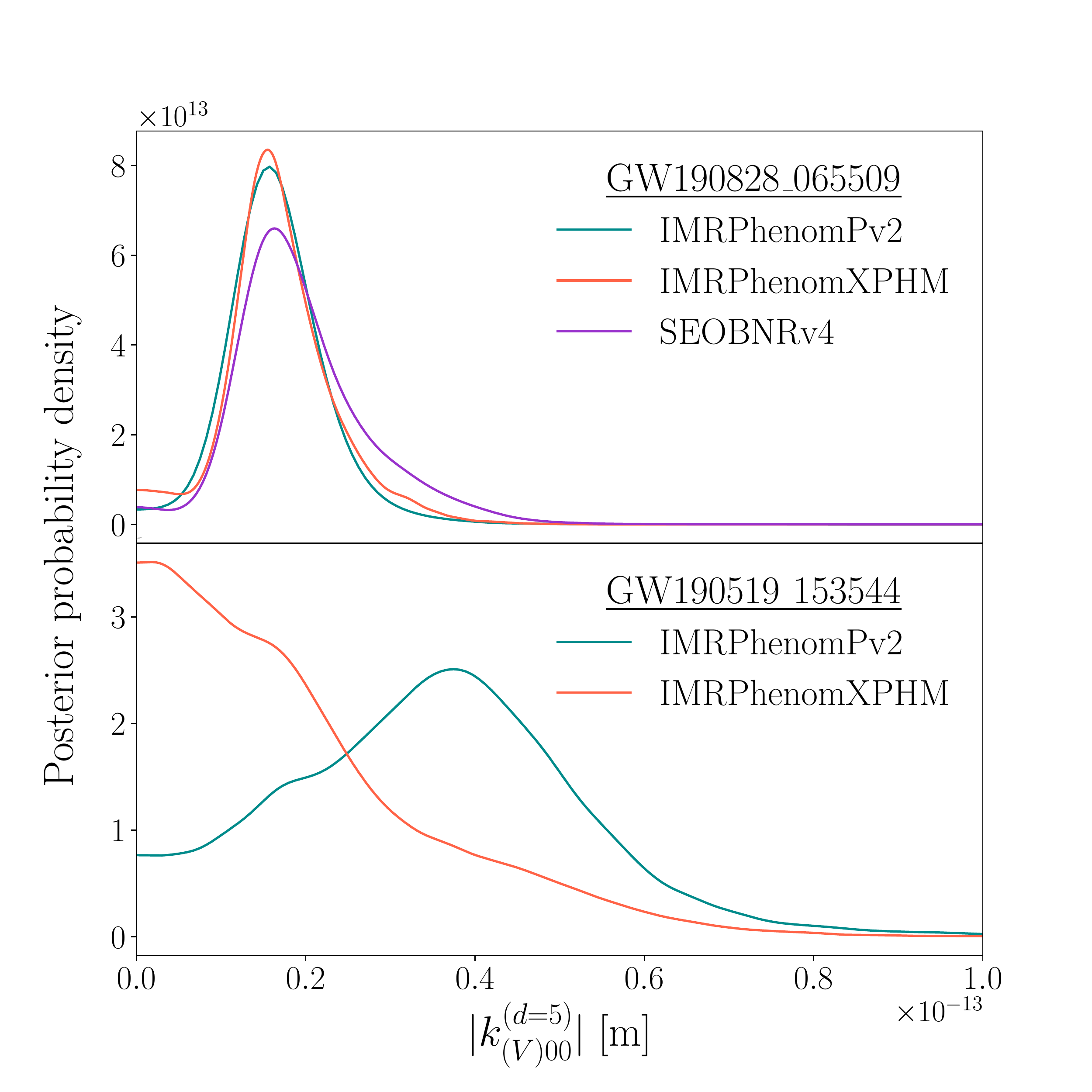}
\caption{\label{fig:kv500_approx} Posterior probability on the isotropic dispersion coefficient $|k_{(V)00}^{(5)}|$ obtained with different waveform models. The top figure presents consistent estimation while the bottom figure presents a case where the probability shape is different according to the waveform model used for inference. }
\end{figure}

In order to investigate the robustness of the results,  we inferred $|k_{(V)00}^{(5)}|$ using different waveform models for several binary black hole events,  as shown for two cases on Fig.~\ref{fig:kv500_approx}.
We compared the posterior probability densities obtained with \texttt{IMRPhenomPv2} with the ones inferred with the \texttt{SEOBNRv4},  that uses an effective one-body description of the dynamics of spinning binaries~\cite{Bohe:2016gbl}; and \texttt{IMRPhenomXPHM},  an updated version of \texttt{IMRPhenomPv2} including higher harmonics and updated calibration to precession~\cite{Pratten:2020ceb}. 
For the 23 events where we compared \texttt{IMRPhenomPv2} and \texttt{IMRPhenomXPHM},  only four events showed a considerable modification of the credible intervals with a mode different from zero for only one of the model (\texttt{GW190519\_153544},  \texttt{GW190706\_222641},  \texttt{GW200219\_094415},  \texttt{GW200225\_060421}).
For the four events where we compared \texttt{IMRPhenomPv2} and \texttt{SEOBNRv4},  two presented a modification of the credible intervals (\texttt{GW190630\_185205},  \texttt{GW190720\_000836}). 
When investigating the events presenting a \acrshort{ci} not including 0,  only three of the ten events are compatible with the \acrshort{gr} case with another model (\texttt{GW190519\_153544},  \texttt{GW190706\_222641},  \texttt{GW190720\_000836}). 
While this rules out mismodelling as the unique source of tension,  it points towards a lack of dynamics in the underlying model for some cases,  and highlights the sensitivity of our analysis to spot deviations in \acrshort{gw} signals.
However,  as waveform models share some common assumptions,  and the uncertainty due to the modeling process (i.e., their mismatch with numerical relativity simulations) is not propagated during the analysis,  more detailed study about waveform accuracy must be carried before ruling it out as a cause for apparent deviations from \acrshort{gr}.

\begin{figure}[h!]
\includegraphics[width=\linewidth]{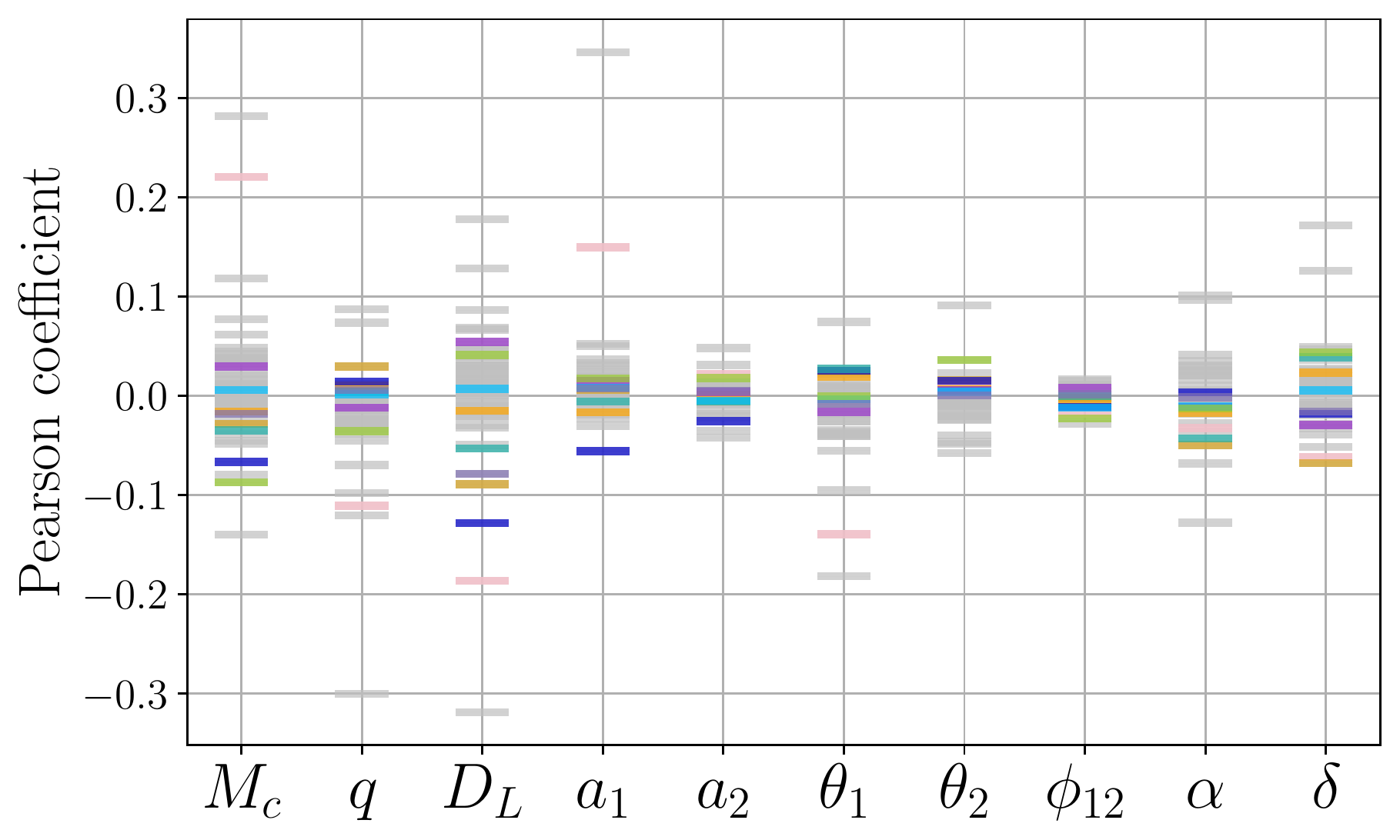}
\caption{\label{fig:corr2} Correlations between $|k_{(V)00}^{(5)}|$ and the source parameters. The x-axis shows the chirp mass $M_c$,  the mass ratio $q$,  the luminosity distance $D_L$,  the spin magnitudes $a_{1}$ and $a_2$,  the spin tilt angles $\theta_{1,  2}$,  the projected angle difference between spins $\phi_{12}$,  the right ascension $\alpha$,  and the declination $\delta$. The colored markers corresponding to the events presenting a deviation on Fig.~\ref{fig:kv500},  with \texttt{GW170814} in pink and \texttt{GW190519\_153544} in blue; the grey markers are the other events. }
\end{figure}

By measuring jointly the source and symmetry breaking parameters,  the correlations between the variables are taken into account during the inference. 
We evaluate them by measuring the Pearson coefficients between $|k_{(V)00}^{(5)}|$ and the source parameters as shown in Fig.~\ref{fig:corr2},  where the events presenting deviations from \acrshort{gr} at 68.3\% \acrshort{ci} are highlighted.
Most events show no or very moderate correlations,  and amongst the highlighted events,  while \texttt{GW170814} and \texttt{GW190519\_153544} present large (anti)correlation with the mass and spin parameters,  other events in agreement with \acrshort{gr} present larger correlations. 
Those results indicate that a more accurate measurement of the source parameters,  as could be obtained from higher \acrshort{snr} or from the detection of higher modes,  can lead to an improvement of the constraints on the \acrshort{sme} coefficients. 

\section{Discussion}
\label{sec:conclusion}

This work presents a new probe of Lorentz and \acrshort{cpt} violation with \acrshort{gws},  extending the search for possible signals from a unified theory of physics.
Our analysis relies on an effective field theory framework,  which allows one to derive phenomenological consequences of spacetime-symmetry breaking across many regimes.
This work complements existing parameterizations of \acrshort{li} violation in \acrshort{gws}.
We extend the measurement of \acrshort{sme} coefficients to higher mass dimension terms in the action,  compared to constraints derived from speed of gravity tests,  by probing \acrshort{gw} dispersion,  including anisotropic and birefringence effects.
Our method goes beyond existing measurements by performing a joint inference of source parameters and symmetry-breaking coefficients,  effectively taking into account correlations.
Compared to \acrshort{sme} measurements relying on time delay measurements,  estimated from posterior probabilities inferred assuming \acrshort{gr},  we find milder constraints on the order of $\mathcal{O}(10^{-13} \ \text{m})$ instead of $\mathcal{O}(10^{-16} \ \text{m})$~\cite{shao20,  wang21}. 
These results indicate that correlations between \acrshort{gr} and \acrshort{sme} coefficients must be taken into account and the simplified treatments in earlier work should be replaced with proper parameter estimation.

While this work was carried,  another team performed an independent measurement of \acrshort{sme} coefficients for $d=5$~\cite{Niu:2022yhr}.
Our analysis differs by estimating the joint posterior probability for 16 $k_{(V)ij}^{(d=5)}$ coefficients,  while they perform a measurement of single and dual coefficients only,  assuming the remaining coefficients to be zero.
Consequently,  our analysis includes possible correlations between \acrshort{sme} coefficients,  as can be seen in Fig.~\ref{fig:kv5ij}.
On the methodological side,  our analysis relies on the \texttt{LALInference} software,  while~\cite{Niu:2022yhr} relies on the \texttt{bilby} software.
Using different methods enables us to verify the validity of the inferences provided by each software,  and we find our results to be in agreement as we both derive a joint constraint around $|k_{(V)00}^{(d=5)}| < \mathcal{O} (10^{-15} \ \text{m})$ when assuming that all other coefficients are zero.

Note also that the main measurement results in this paper,  the limits on coefficients in Table \ref{tab:kij_svd},  can be directly compared to the results from other tests in gravity~\cite{datatables}.
For example,  solar-system tests like lunar laser ranging have yielded measurements on $15$ linear combinations of the $60$ independent mass dimension $5$ coefficients in $(q^{(5)})^{\al\be\ga\mu\nu\rh\si}$ in Eq.~\rf{gravlag},  with limits on the order of $10^3$ m ~\cite{Bourgoin:2020ckq}.
This seems substantially poorer than the limits in this paper from \acrshort{gws},  but the $16$ $k_{(V)ij}^{(d=5)}$ coefficients probed in this paper are distinct linear combinations of the $(q^{(5)})^{\al\be\ga\mu\nu\rh\si}$ coefficients from those occurring in lunar laser ranging.
Similarly,  limits from pulsars via orbital tests are on the order of $10^6 $ m ~\cite{Shao:2018vul},  but probe distinct coefficients from \acrshort{gws}.
Should a nonzero detection occur,  it will be important to compare measurements in distinct tests.

While our global results are compatible with \acrshort{gr},  a subset of events show non-zero $k_{(V)00}^{(d=5)}$ estimates.
We investigated possible shared features between those events,  
that do not display similar sky localisation neither other common parameters.
Robustness tests indicate that for few events,  the addition of higher modes resolves the tension.
However,  several events do not present modified posterior probability profiles when using other waveform models,  pointing to the possibility that existing waveform models may lack dynamical features degenerate with the effects of dispersion.
The current efforts in creating more accurate waveform templates,  e.g. with  the addition of eccentric trajectories,  will provide a better understanding of the relevance of modelling accuracy for \acrshort{sme} tests in particular and tests of \acrshort{gr} in general.

The analysis in this paper studies propagation effects from \acrshort{li} violation.
The addition of other possible effects from higher-order (in $h_\mn$) terms in the \acrshort{sme} on the waveform itself,  
for example,  via a post-Newtonian multipole expansion in the \acrshort{sme} framework,  could provide new tests.
The latter work is in progess~\cite{Xu:2021dcw,  b21}.

\begin{acknowledgments}

LH is supported by the Swiss National Science Foundation grant 199307, and by the European Union’s Horizon 2020 research and innovation programme under the Marie Skłodowska-Curie grant agreement No 945298-ParisRegionFP,.
She is a Fellow of Paris Region Fellowship Programme supported by the Paris Region, and acknowledges the support of the COST Action CA18108. 
Work on this project by QGB and KOA was supported by the United States National Science Foundation (NSF) under grant numbers 1806871 and 2207734.
Work on this project by JDT and MB was supported by the NSF under grant number 1806990.
LS was supported by the National Natural Science Foundation of China (11975027,  11991053,  11721303),  the National SKA Program of China (2020SKA0120300),  and the Max Planck Partner Group Program funded by the Max Planck Society.

The authors would like to thank the LIGO-Virgo-KAGRA collaboration for general support,  and particularly Duncan MacLeod,  Peter Tsun Ho Pang,  Charlie Hoy,  Geraint Pratten and John Veitch for the useful discussion concerning the software infrastructure. 
This research has made use of data or software obtained from the Gravitational Wave Open Science Center (gw-openscience.org),  a service of LIGO Laboratory,  the LIGO Scientific Collaboration,  the Virgo Collaboration,  and KAGRA. LIGO Laboratory and Advanced LIGO are funded by the NSF as well as the Science and Technology Facilities Council (STFC) of the United Kingdom,  the Max-Planck-Society (MPS),  and the State of Niedersachsen/Germany for support of the construction of Advanced LIGO and construction and operation of the GEO600 detector. Additional support for Advanced LIGO was provided by the Australian Research Council. Virgo is funded,  through the European Gravitational Observatory (EGO),  by the French Centre National de Recherche Scientifique (CNRS),  the Italian Istituto Nazionale di Fisica Nucleare (INFN) and the Dutch Nikhef,  with contributions by institutions from Belgium,  Germany,  Greece,  Hungary,  Ireland,  Japan,  Monaco,  Poland,  Portugal,  Spain. The construction and operation of KAGRA are funded by Ministry of Education,  Culture,  Sports,  Science and Technology (MEXT),  and Japan Society for the Promotion of Science (JSPS),  National Research Foundation (NRF) and Ministry of Science and ICT (MSIT) in Korea,  Academia Sinica (AS) and the Ministry of Science and Technology (MoST) in Taiwan.

The authors are grateful for computational resources provided by the LIGO Laboratory and supported by the NSF PHY-0757058 and PHY-0823459.
\end{acknowledgments}

\bibliography{ssb_gwtc3}% Produces the bibliography via BibTeX.

\end{document}